\documentclass[twocolumn]{aastex702}
\usepackage{acronym}
\usepackage{amsmath, amssymb}
\usepackage[dvipsnames]{xcolor}
\usepackage{enumitem}

\newcommand{\bavg}{B_{\rm avg}}
\newcommand{\fluxfs}{\Phi_{\rm FS}}
\newcommand{\euv}{EUV\,304\,\AA}

\newcommand{\afthfs}{\acs{aft}+HFS}

\graphicspath{{./}{figures/}}

\begin{document}
\acrodef{sftm}[SFTM]{Surface Flux Transport Model}
\acrodef{sft}[SFT]{Surface Flux Transport}
\acrodef{hfam}[FAM]{Far-side Acoustic Map}
\acrodef{pinn}[PINN]{Physics-Informed Neural Network}
\acrodef{ml}[ML]{machine learning}
\acrodef{cme}[CME]{coronal mass ejection}
\acrodef{ar}[AR]{active region}
\acrodef{nrt}[NRT]{near-real-time}
\acrodef{sdo}[SDO]{Solar Dynamics Observatory}
\acrodef{hmi}[HMI]{Helioseismic and Magnetic Imager}
\acrodef{phi}[PHI]{Polarimetric and Helioseismic Imager}
\acrodef{so}[SO]{Solar Orbiter}
\acrodef{stereo}[STEREO]{Solar Terrestrial Relations Observatory}
\acrodef{euvi}[EUVI]{Extreme Ultraviolet Imager}
\acrodef{euv}[EUV]{extreme ultraviolet }
\acrodef{gong}[GONG]{Global Oscillation Network Group}
\acrodef{los}[LOS]{line-of-sight}
\acrodef{jsoc}[JSOC]{Joint Science Operations Center}
\acrodef{cr}[CR]{Carrington Rotation}
\acrodef{urm}[URM]{under-represented minorities}
\acrodef{aftm}[AFT]{Advective Flux Transport}
\acrodef{adapt}[ADAPT]{Air Force Data Assimilative Photospheric Flux Transport}
\acrodef{oft}[OFT]{Open Flux Transport}
\acrodef{hipft}[HipFT]{High-Performance Flux Transport}
\acrodef{reu}[REU]{Research Experience for Undergraduate students}
\acrodef{pmi}[PMI]{Photospheric Magnetic field Imager}
\acrodef{pde}[PDE]{Partial Differential Equation}
\acrodef{fno}[FNO]{Fourier Neural Operator}
\acrodef{dnn}[DNN]{Deep Neural Network}
\acrodef{cnn}[CNN]{Convolutional Neural Networks}
\acrodef{aft}[AFT]{Advective Flux Transport}
\acrodef{dft}[DFT]{discrete Fourier transform}
\acrodef{ace}[ACE]{Advanced Composition Explorer}
\acrodef{dscovr}[DSCOVR]{Deep Space Climate Observatory}
\acrodef{hifar}[HIFAR]{helioseismic images of far-side active regions}
\acrodef{hmifsmag}[HMI-FSMAG]{HMI Far-Side Magnetic Maps inferred from helioseismic phase-shift observations}
\acrodef{aia}[AIA]{Atmospheric Imaging Assembly}
\acrodef{nso}[NSO]{National Solar Observatory}
\acrodef{mle}[MLE]{maximum likelihood estimation}
\acrodef{auc}[AUC]{area under the curve}
\acrodef{sse}[SSE]{standard statistical error}
\acrodef{roc}[ROC]{receiver operating characteristic}
\acrodef{auc}[AUC]{area under the curve}
\acrodef{farm}[FARM]{Far-side Active Region Model}
\acrodef{pfss}[PFSS]{potential field source surface}
\acrodef{lff}[LFF]{linear force free}
\acrodef{nlff}[NLFF]{non-linear force free}
\acrodef{rocss}[ROCSS]{relative operating characteristic skill score}
\acrodef{cnn}[CNN]{convolutional neural network}

\newcommand{\fsmag}{\acs{hifar}M}
\newcommand{\hmifs}{\ac{hifar}}
\newcommand{\euvi}{\ac{stereo}/\ac{euvi} 304\,\AA}
\newcommand{\pmodel}{probabilistic \ac{ar} detection model}
\newcommand{\phihifar}{\Phi_\mathrm{HIFARM}}

\title{A Probabilistic Framework for Incorporating Helioseismic Far-Side Active Regions into SFT Models}

\author[0000-0003-3191-4625]{Bibhuti Kumar Jha}
\affiliation{Southwest Research Institute, Boulder, CO 80302, USA}
\email{bkjha.sun@gmail.com}

\author[0000-0003-0621-4803]{Lisa A. Upton}
\affiliation{Southwest Research Institute, Boulder, CO 80302, USA}
\email{maitraibibhu@gmail.com}

\author[0000-0003-0026-931X]{K. D. Leka}
\affiliation{Northwest Research Associates, 3380 Mitchell Lane, Boulder, CO 80301, USA}
\email{leka@nwra.com}

\author[0000-0002-2632-130X]{Ruizhu Chen}
\affiliation{W. W. Hansen Experimental Physics Laboratory, Stanford University, Stanford, CA 94305-4085, USA}
\email{rzchen@stanford.edu}

\author[0000-0002-6308-872X]{Junwei Zhao}
\affiliation{W. W. Hansen Experimental Physics Laboratory, Stanford University, Stanford, CA 94305-4085, USA}
\email{junwei@sun.stanford.edu}

\author[0000-0002-3631-6491]{Shea Hess Webber}
\affiliation{W. W. Hansen Experimental Physics Laboratory, Stanford University, Stanford, CA 94305-4085, USA}
\email{shessweb@stanford.edu}

\author[0000-0001-5503-0491]{Ignacio Ugarte-Urra}
\affiliation{Space Science Division, Naval Research Laboratory, Washington, DC, USA;}
\email{ignacio.ugarteurra.civ@us.navy.mil}

\author[0000-0002-1905-1639]{Kiran Jain}
\affiliation{National Solar Observatory, 3665 Discovery Dr., Boulder, CO 80303, USA}
\email{maitraibibhu@gmail.com}

\shorttitle{Framework for Incorporating HIFAR into SFT Model}
\shortauthors{B. K. Jha et al.}
\correspondingauthor{Bibhuti Kumar Jha}


\begin{abstract}

Surface Flux Transport (SFT) models are routinely used to model the Sun's photospheric magnetic field and provide inner boundary conditions for coronal and heliospheric models, yet they remain fundamentally limited by the lack of direct information about flux emergence on the far side of the Sun. To address this, we develop a framework for incorporating helioseismic images of far-side active regions (HIFAR), into the \acf{aft} model. Using near-simultaneous magnetic flux maps (HIFARM) inferred from HIFAR, and STEREO/EUVI 304\,\AA\ observations from 2010 May 13 to 2014 August 18, we develop a logistic regression model to estimate the probability that a helioseismic detection corresponds to an \acs{ar} based on descriptors of magnetic flux, field strength, and location.  
The model achieves a precision of 0.93 for \ac{ar} detections at a probability threshold of 0.73, selected to reject 90\% of ``ghost'' \acp{ar}. We also derive an empirical scaling relationship between \fsmag\ and \acs{aft} flux to place \fsmag\ on the \acs{aft} flux scale and demonstrate that the framework reproduces the flux evolution of \acp{ar} consistent with near-side observations over multiple solar rotations. At the global scale, we find that conventional near-side-driven \acs{aft} simulations underestimate the total unsigned magnetic flux by $\approx 10-20\%$ of the Sun's total magnetic flux budget. The probabilistic \ac{ar} detection model and the complete framework developed in this work provide a practical pathway for incorporating helioseismic far-side \acp{ar} into SFT models, a step toward realistic modeling of full-Sun magnetic field.
\end{abstract}


\section{Introduction}\label{sec:intro}
\acresetall


The Sun's photospheric magnetic field serves as the necessary lower boundary condition for heliospheric and coronal models \citep{Arge2000,Schrijver2003, Downs2025}, and thus plays a central role in understanding and forecasting space weather. However, routine observations of the Sun are predominantly obtained from a single vantage point, limiting our visibility  to only one hemisphere of the Sun at any given time, referred to as the visible or near-side of the Sun. Historically, this observational limitation has been addressed ``synoptically'', constructing full-Sun magnetic maps by combining observations taken over one complete solar rotation ($\sim27.3$\,days) to form what are known as Carrington maps \citep{Harvey1998, Ulrich2006, Sheeley2011}. This approach can characterize the large-scale structure of the heliosphere; however, merging approximately a month-long set of data to create a composite inherently assumes that the global magnetic configuration evolves slowly over a rotation. As a result, synoptic maps provide a static representation of the solar magnetic field and can not capture any dynamic emergence or evolution of magnetic structures on timescales smaller than $\sim1$\,month.

Over the past few decades, \ac{sft} models have been developed to reconstruct the full-Sun photospheric magnetic field \citep{DeVore1984, Worden2000,Schrijver2003, Sheeley2005, Upton2014, Hickmann2015, Jiang2014a, Yeates2023, Caplan2025}. \ac{sft} models attempt to reconstruct the full-Sun photospheric magnetic field by evolving the near-side observation, through the combined effects of advection and diffusion, driven by large-scale axisymmetric flows and non-axisymmetric turbulent convective flows \citep{Upton2014, Upton2014a, Yeates2023}.

\ac{sft} models have proven to be highly effective for space-weather modeling and forecasting by providing an observationally constrained representation of the photospheric magnetic field for coronal models, e.g. \acl{pfss} \citep[PFSS,][]{Schatten1969, Altschuler1969, Wang1992}. However, because \ac{sft} models are primarily driven by near-side observations, they do not capture newly emerging \acp{ar} on the far-side of the Sun or account for additional flux in regions that continue to grow during their far-side transit.

\begin{figure}[t!]
    \centering
    \includegraphics[width=\columnwidth]{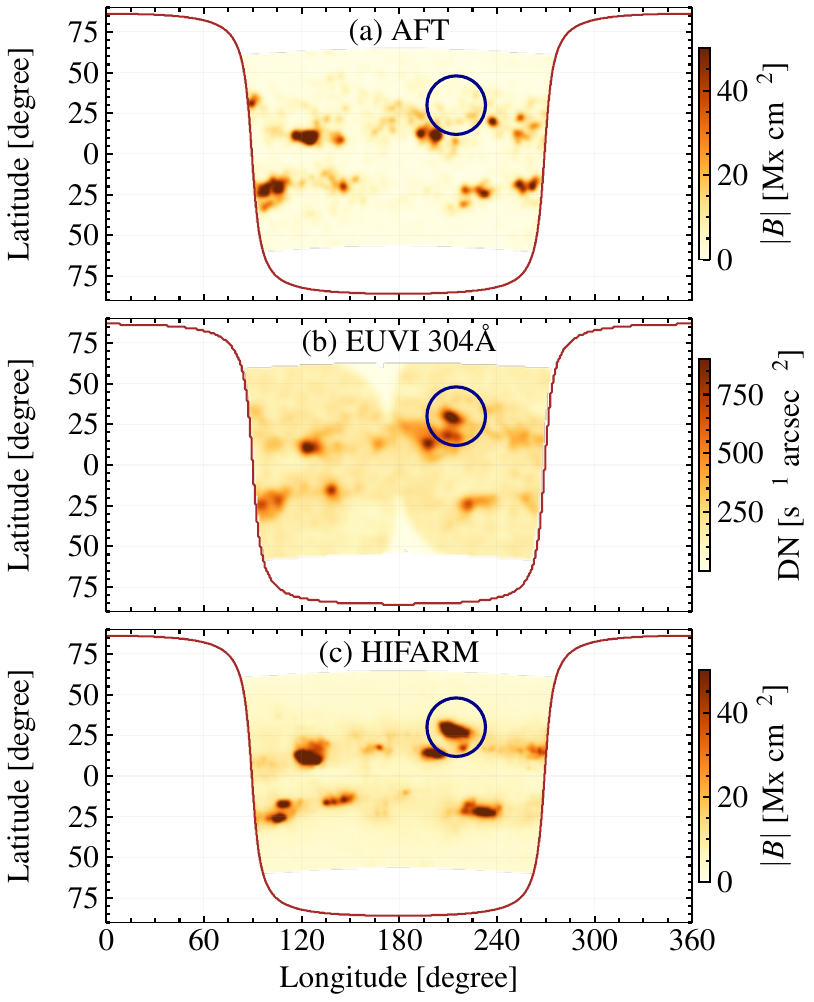}
    \caption{Far-side representations  
    of the Sun in heliographic Carrington grid, from (a) an \acs{aft} \citep{Upton2014} model, (b) \euvi\ observations, and (c) \acs{hifar}, from 2012 January 10. The circle marks an \acp{ar} that emerged and evolved on the far side of the Sun but are completely missed in the modeled representation, highlighting the current limitations of near-side-driven \ac{sft} models in reconstructing the full-sun photospheric magnetic field.}
    \label{fig:allfs}
\end{figure}

In \autoref{fig:allfs}, we highlight this fundamental limitation of conventional \ac{sft} models. As an example, the \ac{ar} (marked with a circle) in \autoref{fig:allfs} is entirely absent from the far-side magnetic field representation produced by the \ac{sft} model (\autoref{fig:allfs}a), yet it is clearly visible in observations from \aclu{euvi} \citep[EUVI,][]{Howard2008} on board \aclu{stereo} \citep[STEREO; \autoref{fig:allfs}b][]{Driesman2008}. Although such missing or misrepresented \acp{ar} do not significantly affect long-term studies of the Sun's global magnetic field evolution \citep{Schrijver2001, Jiang2013, Virtanen2021, Jha2026}, they introduce substantial inaccuracies into the far-side surface magnetic field maps. Consequently, coronal and heliospheric models driven by these maps produce results that are inconsistent with the reality.

\citet{Arge2013}, and more recently \citet{Heinemann2025b}, reported that incorporating far-side \acp{ar} into \ac{sft} significantly improves solar-wind forecasting, in contrast to the findings of \citet{Wang2024} and \citet{Knizhnik2024}, who found little to no effect on coronal and heliospheric modeling. Resolving these conflicting results requires more accurate representations of the far-side magnetic field. Therefore, incorporating improved far-side magnetic field information into \ac{sft} models is a promising pathway toward reducing these uncertainties and improving space-weather forecasting capabilities \citep{Arge2013, Leka2026}.

In a recent work, \euv\ observations were used as a proxy for the strength of magnetic field on the far side \citep{UgarteUrra2015} and incorporated into the \ac{aft} model to produce far side informed photospheric magnetic maps \citep[][`AFT+304' maps hereafter]{Upton2024}. Their results emphasize that accurately capturing the emergence and growth of \acp{ar} on the far-side is essential, as missing or misrepresented regions can represent a significant fraction of the flux on the far-side of the Sun. Unfortunately, \euv\ observations of the Sun's far side are not continuously available, due to the \ac{stereo} orbit and the loss of one of the spacecraft, limiting the ability of these direct observations to inform \ac{sft} models.

Advances in helioseismology \citep{Gizon2005} have made it possible to monitor \acp{ar} beyond direct view by inferring far-side solar activity from near-side observations \citep{GonzalezHernandez2007, Lindsey2017a, Zhao2019, Chen2022}. Currently, the \aclu{nso}/\aclu{gong} \citep[NSO/GONG,][]{Harvey1996, Jain2026} and \aclu{sdo}/\aclu{hmi} \citep[SDO/HMI,][]{Pesnell2012, Schou1998} provide regular helioseismic analysis of far-side acoustic travel-times in the form of phase-shift maps \citep{Lindsey2000, Zhao2019}. These phase-shift maps exhibit a strong correlation with the presence and distribution of the magnetic field \citep{Lindsey1990, Lindsey2000, Lindsey2017a, Chen2022, Hamada2026}, providing an indirect means of estimating magnetic activity on the far side of the Sun. 

However, the physical relationship between the inferred phase shifts and the underlying magnetic field remains poorly constrained due to the complex nature of wave propagation through magnetized solar plasma \citep{GonzalezHernandez2007, Lindsey2017a}. Consequently, a variety of empirical \citep{GonzalezHernandez2007, Yang2024, Hamada2026} and machine-learning-based \citep{Chen2022} approaches have been developed to translate helioseismic phase-shift maps into estimates of far-side magnetic flux distributions. Figure~\ref{fig:allfs}c shows an example of a far-side magnetic flux density map derived by \citet{Chen2022} and its close correspondence with \euv\ observations, making helioseismic far-side maps a promising alternative to \euv\ observations for incorporating systematic far-side information into \ac{sft} models \citep{Jain2023}.

Far-side helioseismology suffers from several sources of uncertainty and inconsistency. In particular, phase-shift maps are known to include significant noise, have limited spatial resolution, and can produce `ghost \acp{ar}' \citep[false signals;][]{Zhao2019, Hamada2025}. The precise origin of these ghosts \acp{ar} is not yet fully understood; however, it is widely suspected that the depth and effective spatial resolution of far-side helioseismic measurements, combined with projection effects and strong helioseismic center-to-limb effect \citep{Chen2018}, contribute significantly to the uncertainty in both the detection and localization of far-side magnetic activity. \citet{liewer2012,Liewer2014} reported that approximately $10\%$ of \acp{ar} identified in far-side helioseismic observations from \ac{gong} exhibit no corresponding intensity brightening in \euvi\ observations and therefore appear as ghost \acp{ar}. Consequently, careful evaluation of helioseismic far-side detections is required before incorporating them into an \ac{sft} model.

In this article, we focus on characterizing, mitigating, and minimizing the uncertainties inherent in the detection of far-side \acp{ar}, including the presence of ghost \acp{ar} and uncertainties in their locations and magnetic flux content. We begin by introducing a statistically informed \ac{ar} detection technique designed to identify far-side activity while minimizing the possibilities of including far-side helioseismic artifacts. Building upon this, we exploit the maximum available information from the detected far-side \acp{ar} and incorporate it into the \ac{sft} model, with the goal of improving the realism and accuracy of modeled photospheric magnetic field evolution, thereby improving the boundary conditions used in heliospheric and coronal modeling.

\section{Advective Flux Transport (AFT) Model}\label{sec:aft}

The \ac{aft} model is an \ac{sft} model that solves the radial component of the magnetic transport equation \citep[see Equation~1 in][]{Upton2014} to evolve photospheric magnetic flux under the influence of horizontal surface flows \citep{Upton2014, Upton2014a}. In \ac{aft}, these surface flows include axisymmetric components, namely differential rotation and meridional flow, as well as non-axisymmetric supergranular motions \citep{Hathaway2011, RightmireUpton2012, Upton2014, Upton2014a}. The model regularly assimilates line-of-sight (LOS) magnetogram observations (e.g., SDO/HMI), corrected with the assumption that all field is radial, to produce an observationally constrained full-Sun photospheric radial magnetic field map. \ac{aft} produces these maps in heliographic Carrington co-ordinate (with grid size $512\times1024$ pixels) with resolution of $0.35\degr\,\mathrm{pixel}^{-1}$ in latitude and longitude.

The capability of \ac{aft} is not restricted to simulating near-side observations; it can also be used for predicting future global magnetic field evolution \citep{Upton2014, Jha2024a} and for historical reconstructions of the photospheric magnetic field \citep{Jha2026}. However, in this study, we use \ac{aft} exclusively in its data assimilation (baseline) mode and will integrate the far-side information to minimize the uncertainty in the full-sun photospheric magnetic map due to missing far-side information.

\section{Far-side Observations}\label{sec:data}

In this study, we compare far-side observations and data products from three complementary sources: (i) \acs{sdo}/\acs{hmi} time-distance \ac{hifar} maps, 
(ii) \euvi\ observations, and (iii) AFT+304 maps from 2010 May 13 to 2014 August 18, when \ac{stereo} provided near-continuous coverage of the far side of the Sun. Together, these datasets provide the observational and model constraints required to identify, characterize, and incorporate helioseismic far-side \acp{ar} data into the \ac{aft} framework and develop the probabilistic \ac{ar} detection method.

\begin{enumerate}[left=-10pt, label={(\textit{\roman*})}]

\item Doppler-velocity observations from \ac{hmi} are used to produce the \hmifs\ data set with a cadence of 12\,hours \citep{Zhao2019}. \citet{Chen2022} used these \hmifs\ data to derive unsigned magnetic-flux density maps using a two-stage \acl{ml} technique. In the first stage, a \acl{cnn} is trained to convert \euvi\ observations into magnetic-flux maps. In the second stage, the relationship between \hmifs\ and EUV-derived magnetic maps is learned. This relationship enables the direct conversion of \hmifs\ into magnetic-flux estimates (\fsmag).

\item Full-Sun He~\textsc{ii} 304\,\AA\ EUV maps are constructed by combining near-simultaneous observations from SDO/AIA and STEREO/EUVI (A and B) to produce continuous 360$^\circ$ maps of the Sun \citep{UgarteUrra2015, Upton2024}. The two instruments are cross-calibrated using a time-dependent correction factor to match the \ac{euvi} intensity scale and mitigate the effects of instrumental degradation. In this work, we use only the far-side portion of these composite maps. This choice ensures consistency with the AFT+304 simulations, which are constructed from the same processed \euv\ dataset.

\item AFT+304 photospheric magnetic field maps \citep{Upton2024} assimilate near-side magnetic flux from \ac{hmi} magnetograms while \acp{ar} on the far side are incorporated using composite EUV 304\,\AA\ intensity maps as a proxy for magnetic activity inferred through a magnetic flux- \euv\ -luminosity relationship \citep{Schrijver1987,UgarteUrra2015}. While this hybrid approach provides one of the most realistic representations currently available of the evolving full-Sun photospheric magnetic field, it is limited by the availability of \ac{stereo} observations. 
\end{enumerate}

We use \fsmag\ as the primary far-side data source to develop a framework to incorporate helioseismic far-side \acp{ar} into \ac{aft}. The \euv\ observations are used to label the \acp{ar} identified in the \fsmag\ data as `True \ac{ar}' or `ghost \ac{ar}', and to develop the probabilistic model for \ac{ar} detection. In addition, we compare the total unsigned magnetic flux of \acp{ar} identified in \fsmag\ with the corresponding flux in the AFT+304 simulations to establish an empirical relationship between the two. This relationship enables us to estimate the amount of magnetic flux that must be incorporated from \fsmag\ into \ac{aft} to accurately reproduce the observed evolution of \ac{ar} flux.

\section{Methodology}
\subsection{Active Region Detection} \label{subsec:ar_detection}

The detection of \acp{ar} from \fsmag\ and \euv\ far-side maps is performed using an automated adaptive intensity thresholding. To identify significant solar structures, we first must robustly identify the quiescent background.  To do this, we invoke the iterative sigma-clipping\footnote{ \citet{astropy},  \url{https://docs.astropy.org/en/stable/api/astropy.stats.sigma_clipped_stats.html}} method to estimate the background mean ($\mu$) and standard deviation ($\sigma$). This iterative process involves an initial calculation of $\mu$ and $\sigma$, followed by selecting only data points in $\left[\mu-3\sigma, \mu+3\sigma\right]$ and rejecting data outside of this interval. The statistics, $\mu$ and $\sigma$, are then re-computed only from the selected data points iteratively until the values converge (i.e., iterate until the last iteration rejects nothing). 

\begin{figure}
    \centering
    \includegraphics[width=0.9\columnwidth]{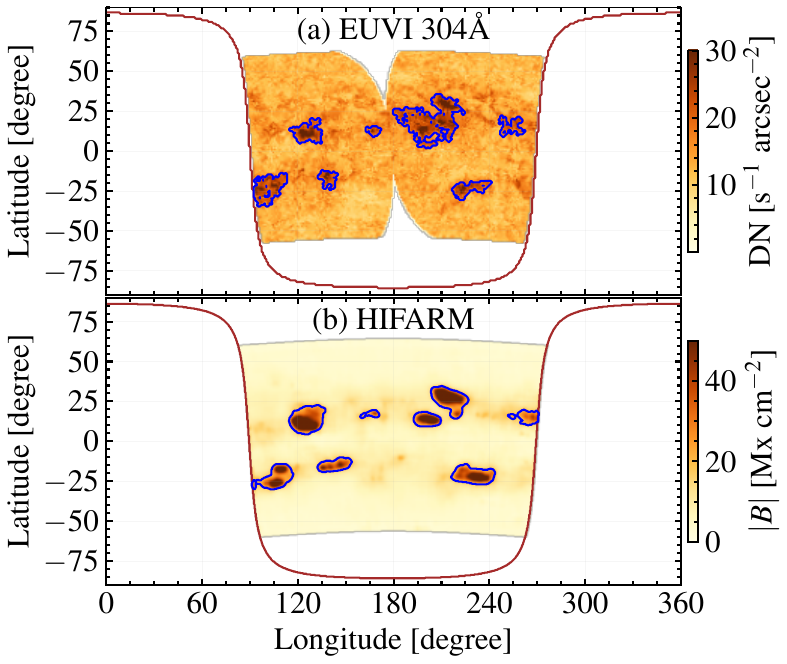}
    \caption{Representative example of \ac{ar} detection in (a) \euv\ observations and (b) \fsmag\ using the adaptive intensity-threshold based on sigma-clipping method, from 2012 January 10. The detected regions are outlined by contours indicating the identified \acp{ar}' boundaries.}
    \label{fig:ardetection}
\end{figure}

The primary benefit of using iterative sigma-clipping over a global mean is its resistance to outliers. In \fsmag\ and \euv\ data, \acp{ar} have high-intensity outliers that can inflate the global $\mu$ and $\sigma$, leading to an overestimated background and potential under-detection of weaker features. By iteratively removing these bright structures from the calculation, the algorithm provides a stable background estimate (against other instrumental and solar cycle phase related fluctuation) for adaptive thresholding, ensuring that the final threshold $T := \mu + k\sigma$ is sensitive enough to isolate \acp{ar}. Here, $k$ is a tuning parameter found to be $k=5$ for \fsmag\ and $k=3$ for \euv. A representative example of the detected \acp{ar} from \euv\ and \fsmag\ are shown in \autoref{fig:ardetection}

We generate a binary mask such that pixels are flagged if their intensity is greater than the optimum threshold. This is followed by a minimal area threshold (in pixels, approximately $120$\,MSH -- millionths of a solar hemisphere, on average, chosen based on visual inspection), to eliminate smaller structures, which are unlikely to be an \ac{ar}. Once all candidate \acp{ar} are isolated, we extract physical parameters. 

For \fsmag\ these parameters are flux-density-weighted heliographic latitude ($\lambda$) and Carrington longitude ($\phi$); total absolute flux ($\Phi$) and average field strength ($B_\mathrm{avg}$) in an \ac{ar}  (see section~\ref{sec:probclass} for discussion). We also calculate the projection factor, $\cos\theta$, defined as the cosine of the angle between the surface normal and the line of sight to the observer ($\theta$), see \citep{Thompson2006} for details. In the near-side, $\cos\theta>0$, where as in the far-side $\cos\theta<0$, with $\cos\theta = -1.0$ at the far-side view's disk center, and $|\cos\theta| < 0.1$ is close to the limb from either near-side or far-side viewpoints.

\subsection{Labeling HIFARM with EUVI 304\,\AA}\label{subsec:ar_labeling}
To distinguish between true \acp{ar} and ghost \acp{ar}, we cross-validate each candidate \ac{ar} identified in the \fsmag\ maps against the corresponding \euv\ observations. For each \fsmag\ detection, we define a bounding box centered on the heliographic coordinates of the region, $\lambda\pm5^\circ$ and $\phi\pm10^\circ$, to account for the uncertainty associated with the location of \acp{ar} in these data. If $\mathcal{M}_A$ and $\mathcal{M}_B$ represent the sets of foreground pixels in the \ac{ar} masks of the \fsmag\ and \euv\ observations, respectively, we classify a helioseismic \ac{ar} as a true \ac{ar} if it overlaps with an \euv\ detected region within this box, i.e., $|\mathcal{M}_A \cap \mathcal{M}_B| > 0$, and as a ghost \ac{ar} otherwise, i.e., $|\mathcal{M}_A \cap \mathcal{M}_B| = 0$. In cases where that part of the Sun is missing an \euv\ map, we label the detection as `unknown.' This statistical approach specifically targets the reliability of \fsmag; we do not include in the analysis \acp{ar} present in \euv\ that lack a helioseismic counterpart. The resulting classification produces a labeled dataset suitable for the statistical modeling and validation of \fsmag.

\subsection{Probabilistic Classification Model}
\label{sec:probclass}
We develop a logistic regression model to estimate the probability that a candidate detection on the far-side is a true \ac{ar}. This approach models the relationship between a binary dependent variable $Y\in\{0,1\}$, and a vector of physical parameters $\mathbf{x}$. Here, $Y=1$ denotes a true \ac{ar} (success), whereas $Y=0$ denotes a ghost \ac{ar} (failure). In this framework, probability of success, $P$ is given by,
\begin{equation}
P(Y=1 \mid \mathbf{x}) := \frac{1}{1 + e^{-L}},
\label{eq:regressionmodel}
\end{equation}
where $L$ is the linear predictor defined as the weighted sum of the input physical parameters ($\mathbf{x}$), and is written as
\begin{equation*}
L = \sum_{i=0}^{N} \beta_i x_i = \mathbf{\beta}^T \mathbf{x}, \quad x_0=1.
\end{equation*}
In this formulation, $\beta_0$, intercept term for $L$ determines the baseline probability of success. The remaining coefficients $\beta_i$ ($0<i\le N$), represent the weights of each of the physical parameters of the \ac{ar}, often referred to as predictors in statistics.

For the \fsmag\ dataset, we adopt set of physical parameters, $\mathbf{x}=(\phihifar, \bavg, \cos{\lambda})$
as our set of predictors, based on their physical characteristics and technical constraints of far-side helioseismology. $\phihifar$ is the primary indicator of the correlation between helioseismic phase shifts and the total magnetic flux \citep{Lindsey1990, Lindsey2000}, whereas $\bavg$ measures the magnetic concentration. Together, these parameters help distinguish coherent, large-scale \acp{ar} from diffuse, low-intensity helioseismic artifacts. We also include $\cos\lambda$ to account for the well-known tendency of the \acp{ar} to emerge preferentially within the activity belts spanning approximately $\pm45\degr$ in latitude \citep{Maunder1903, Hathaway2015}. To assess whether additional physical parameters improve the model, we compare logistic regression models in Appendix~\ref{appendix:comparision}, providing the basis for the adopted predictor set. Therefore, the linear model, $L$ in \autoref{eq:regressionmodel}, takes the form of
\begin{align}
    L(\phihifar, \bavg, \lambda) = \beta_0 &+ \beta_1 \Phi_\mathrm{HIFARM,\,std}\nonumber\\
    &+ \beta_2 B_{\rm avg,\,std}\nonumber\\
    &+ \beta_3 (\cos\lambda)_{\text{std}}.
    \label{eq:linearpred}
\end{align}
In \autoref{eq:linearpred}, coefficients $\beta_i$ for $i= 1, 2, 3$, quantify the relative contributions of the standardized variables $\Phi_\mathrm{HIFARM,\,std}$, $B_{\rm avg,\,std}$ and $(\cos\lambda)_{\rm std}$, respectively. We standardize these parameters to bring their respective mean to zero and standard deviation to unity, allowing a direct comparison of the relative influence of the physical parameters through their corresponding $\mathbf{\beta}$.

\subsection{Estimation of Predictors' Coefficients}
We estimate the coefficients $\mathbf{\beta}$ of the logistic probabilistic model using a sample of $N = 8181$ observations ($\approx80\%$ of the total dataset) through \acl{mle} 
implemented via the \texttt{Logit} class in the \texttt{statsmodels} Python package \citep{seabold2010}. The fitted logistic regression model yields a log-likelihood of $-4229.4$, a likelihood-ratio test $p$-value of $<0.001$, and a pseudo-$R^2$ of 0.159. These results indicate that the selected physical parameters contain meaningful information for distinguishing true and ghost \acp{ar}, despite the inherent uncertainties associated with helioseismic far-side observations. In \autoref{tab:predictor_summary} we report the $\mu$ and $\sigma$ used to standardize the physical parameters and the estimated best-fit coefficients with their corresponding \acp{sse}. For $\phihifar$, we note that $\mu < \sigma$, implying a right skewed distribution of $\phihifar$ and hence, $\mu$ is not the best representative of the distribution. However, we retain the current formulation in this work to keep it consistent with other physical parameters.

\begin{deluxetable}{lrrr}[!ht]
\tablecaption{Summary of the logistic regression predictors. The sample ($N=8181$) $\mu$ and $\sigma$, used to standardize the physical parameters are listed, along with the fitted standardized coefficients and their corresponding uncertainty, $\beta \pm \beta_{\rm SSE}$. \label{tab:predictor_summary}} 
\tablehead{
\colhead{Physical Parameters} & \colhead{$\mu \pm \sigma$} & \colhead{$\beta \pm \beta_{\rm SSE}$}n3
}
\startdata
Intercept & --- & $1.33 \pm 0.04$\\
$\phihifar\,(10^{21}$\,Mx) & $6.81 \pm 09.45$ & $1.22 \pm 0.09$\\
$\bavg$\,(G) & $31.36 \pm 16.11$ & $0.64 \pm 0.06$\\
$\cos\lambda$ & $0.95 \pm 00.04$ & $0.07 \pm 0.02$\\
\enddata
\end{deluxetable}

\subsection{Model Performance Evaluation}\label{subsec:model_eval}
We evaluate the performance of the \pmodel\ to predict whether an \hmifs-detected \ac{ar} (via the \fsmag\ maps) is co-occurring with an \euv-detected \ac{ar},
using an independent subset of 2046 labeled \fsmag\ \acp{ar}. For each \ac{ar}, we compute the probability $P$ using \autoref{eq:regressionmodel}. For a given probability threshold, $P_\mathrm{th}$, we define the possible outcomes of the classification as True Positive (TP), False Positive (FP), True Negative (TN), and False Negative (FN):
\begin{align*}
\mathrm{TP} &\iff P \ge P_{\mathrm{th}} \land |\mathcal{M}_A \cap \mathcal{M}_B| > 0, \\
\mathrm{FP} &\iff P \ge P_{\mathrm{th}} \land |\mathcal{M}_A \cap \mathcal{M}_B| = 0, \\
\mathrm{TN} &\iff P < P_{\mathrm{th}} \land |\mathcal{M}_A \cap \mathcal{M}_B| = 0, \\
\mathrm{FN} &\iff P < P_{\mathrm{th}} \land |\mathcal{M}_A \cap \mathcal{M}_B| > 0.
\end{align*}
Distribution corresponding to these classes are also marked in \autoref{fig:logistic_performence}a. Using these definitions, the True Positive Rate (TPR) and False Positive Rate (FPR) are calculated as
\begin{equation*}
    \mathrm{TPR} = \frac{\mathrm{TP}}{\mathrm{TP} + \mathrm{FN}}, \qquad
    \mathrm{FPR} = \frac{\mathrm{FP}}{\mathrm{FP} + \mathrm{TN}}.
\end{equation*}
It must be stressed that all entries require the existence of an \fsmag-detected \ac{ar}, and that the negative class does not represent the lack of an \ac{ar} in \fsmag, but rather and \fsmag\ candidate \ac{ar} that does not satisfy the model criteria. These metrics, either individually or in combination, are commonly used to determine the optimal $P_{\mathrm{th}}$ for a particular scientific objective \citep[see,][and the references therein for details]{Bloomfield2012, Barnes2016, Kubo2019}. In this work, however, we specify a target FPR and determine the corresponding $P_{\mathrm{th}}$.

\subsection{Calibration of HIFARM Flux for AFT}
The magnetic flux inferred from far-side helioseismic maps differs significantly from the flux required to reproduce the observed evolution of \acp{ar} as they rotate onto the visible hemisphere and become directly observable \citep{Leka2026}. Consequently, direct incorporation of \fsmag{s}, without proper scaling, into \ac{aft} can lead to unrealistic \ac{ar} strengths and evolution. Therefore, before incorporating \fsmag{s} into \ac{aft}, the \fsmag\ flux must be appropriately scaled to avoid a discontinuities and jumps in \acp{ar} flux when the regions rotate on to the Earth-facing hemisphere of the Sun.
\citet{Leka2026} adopted a constant scaling factor ($\alpha$) when incorporating helioseismic far-side \acp{ar} detected from \acs{gong} observations into \ac{aft}. However, because the sensitivity of \hmifs\ depends on the magnetic flux distribution, and the relative location of each \ac{ar} with respect to the far-side disk center, $\alpha$ may depend on $\phihifar$, $\bavg$, and $\theta$. Therefore, rather than assuming a constant scaling factor, we derive an empirical relationship between the magnetic flux required by \ac{aft} on the far side, $\fluxfs$ and $\phihifar$. We model this relationship as,


\begin{equation*}
    \fluxfs=\alpha\left(\phihifar,\bavg, \theta\right)\phihifar,
\end{equation*}
which provides a physically motivated conversion between the two flux estimates.

We estimate $\fluxfs$ using the AFT+304 photospheric magnetic field maps, which incorporate far-side \ac{ar} information inferred from \euv\ observations, and therefore assume $\fluxfs \equiv \Phi_{\rm AFT+304}$. We adopt AFT+304 rather than directly using \euv-derived flux estimates because the latter can be overestimated during flaring events, when transient enhancements in \euv\ intensity violate the assumed intensity–flux scaling. Furthermore, continuous assimilation of near-side magnetic field observations enables AFT+304 to correct for uncertainties that accumulate in the photospheric maps due to statistical uncertainties in the \euv\ intensity–magnetic flux scaling relation.

To compute $\Phi_{\rm AFT+304}$, we use the mean heliographic Carrington location, $(\lambda,\phi)$, of true \acp{ar} from the labeled \fsmag\ catalog. We take a bounding box of size $\lambda\pm5\degr$ and $\phi\pm10\degr$ to ensure adequate spatial coverage of each \ac{ar}. We calculate the total unsigned magnetic flux in this box from the AFT+304 maps ($\Phi_{\rm AFT+304}$) and compute the scaling factor, $\alpha=\Phi_{\rm AFT+304}/\phihifar$.
This $\alpha$ is then used to derive the empirical relationship that maps $\left(\phihifar, \bavg, \theta\right)\mapsto\alpha$.

\section{Results}\label{sec:results}
\subsection{Model Performance and Diagnostics}\label{subsec:model_performence_diagnostics}

\begin{figure*}[htbp!]
\centering
\includegraphics[width=\textwidth]{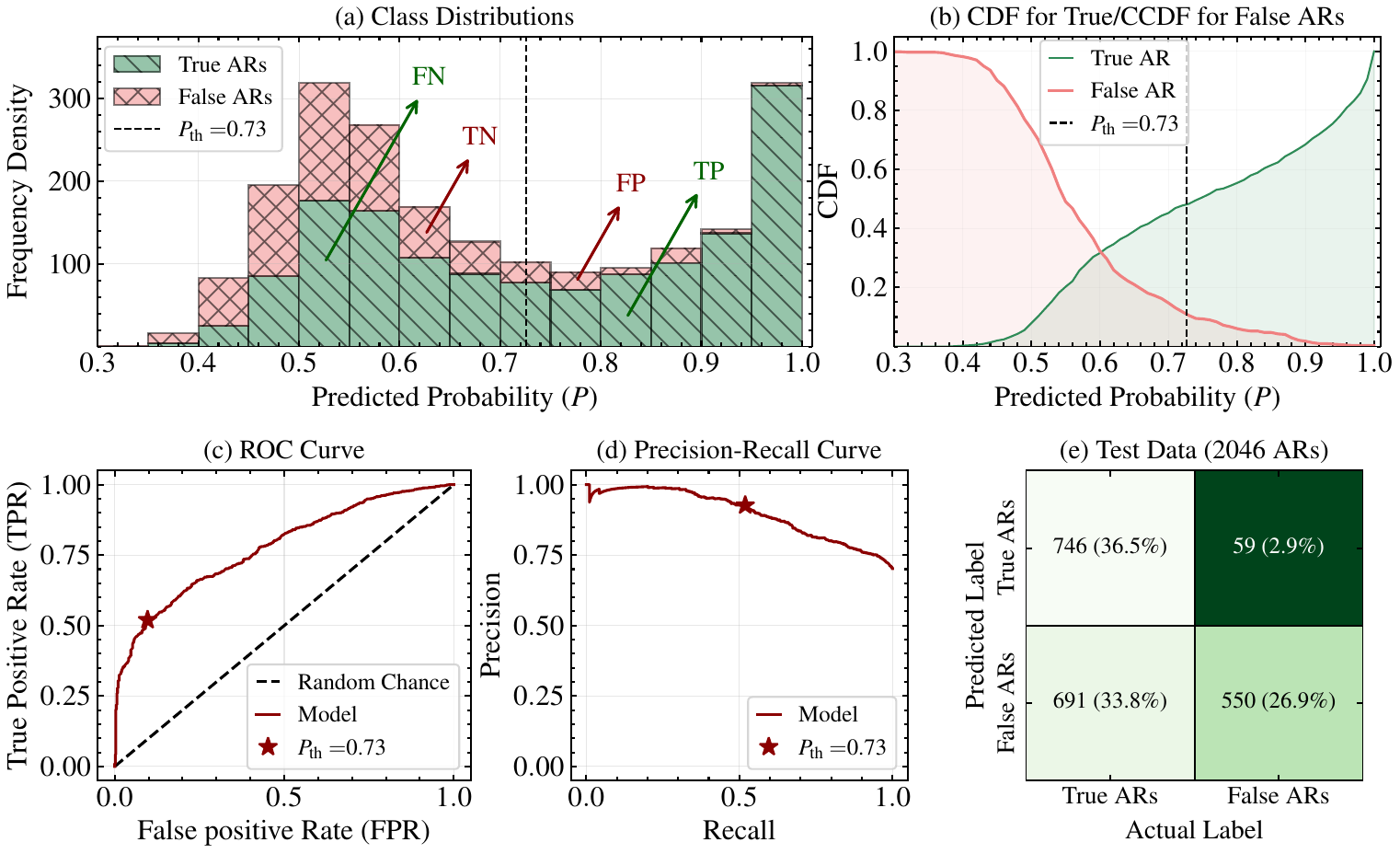}
\caption{Logistic regression model performance metrics for far-side data to be incorporation in \ac{aft}.
(a) Frequency density of $P_\mathrm{th}$ for true ARs (green) and ghost (false) ARs (pink), with $P_\mathrm{th} = 0.73$ indicated (the probability threshold used for classification-based evaluation).
(b) Cumulative distribution functions for both classes, with $P_\mathrm{th} = 0.73$ again indicated.
(c) ROC curve,
(d) Precision-recall (purity-completeness) curve, with $P_\mathrm{th}= 0.73$ indicated by the star.
(e) Confusion matrices evaluated on the test dataset (2046 ARs), showing the model's classification distribution at the $P_{\rm th} = 0.73$ threshold.}
\label{fig:logistic_performence}
\end{figure*}

The performance of the logistic regression model was assessed using an independent subset of $N = 2046$ \acp{ar} (out of 10227), which are not used in estimating the model coefficients $\mathbf{\beta}$, and the results are shown in \autoref{fig:logistic_performence} (a-e). To minimize contamination from false detections, we selected the probability threshold with targeting $\mathrm{FPR}=0.1$ (\autoref{subsec:model_eval}), corresponding to the rejection of approximately 90\% of false \acp{ar} and with $\mathrm{TPR}=0.52$. This criterion yielded a probability threshold of $P_{\rm th}=0.73$. In \autoref{tab:model_performance}, we present the precision, recall, and F1-score, defined in \autoref{tab:metrics_definition}, achieved for true and ghost \acp{ar} with this probability threshold.

\begin{deluxetable}{p{0.02\textwidth} p{0.45\textwidth} @{\hspace{0.1\textwidth}} p{0.35\textwidth}}
\tablecaption{Definition of Classification Metrics
\label{tab:metrics_definition}}

\tablehead{
\colhead{S.N.} & \colhead{Metric Definition} & \colhead{Formula}
}
\startdata
\\
1. & Precision is the fraction of predicted positive instances (i.e., TP) that are actually positive (i.e., True ARs). & $\displaystyle \mathrm{Precision} = \frac{\mathrm{TP}}{\mathrm{TP} + \mathrm{FP}}$\\
2. & Recall, also known as the True Positive Rate (TPR) or Sensitivity, is the fraction of actual positive instances that are correctly identified. & $\displaystyle\mathrm{Recall} = \frac{\mathrm{TP}}{\mathrm{TP} + \mathrm{FN}}$ \\
3. & \Acf{roc} shows the variation of True Positive Rate (TPR) against the False Positive Rate (FPR), as defined in Section~\ref{subsec:ar_labeling}, at various classification thresholds $P_\mathrm{th}$. & 
$\displaystyle
\mathrm{ROC}=\left\{\left(\mathrm{FPR}(P_\mathrm{th}), \mathrm{TPR}(P_\mathrm{th}) \right)\,:\,P_\mathrm{th} \in [0,1]\right\}$\\
4. & \Acf{auc} is the measure of the overall ability of a classifier to distinguish between positive and negative classes. A value of 1 indicates perfect classification and 0.5 corresponds to random guessing.&
$\displaystyle
\mathrm{AUC} = \int_{0}^{1} \mathrm{TPR}(\mathrm{FPR}) \, d(\mathrm{FPR})$\\
5. & \Acf{rocss} is another metric calculated from \ac{auc} and routinely used in statistics. &
$\displaystyle \mathrm{ROCSS}:=2\mathrm{AUC}-1$ \\
6. &  F1-Score is defined as the harmonic mean of precision and recall. &
$\displaystyle
\frac{1}{\text{F1-score}} =
\frac{1}{2}
\left(
\frac{1}{\mathrm{Precision}} +
\frac{1}{\mathrm{Recall}}
\right)$
\rule{0pt}{4ex}
\enddata
\end{deluxetable}

\begin{deluxetable}{lcc}[htbp!]
\tablecaption{
Classification metrics evaluated using an independent subset of $N=2046$ \acp{ar} at the adopted probability threshold, $P_{\rm th}=0.73$. The threshold was selected by targeting a false-detection rate of 0.1, corresponding to the rejection of approximately 90\% of ghost \acp{ar}.
\label{tab:model_performance}
}
\tablehead{
\colhead{Metric} &
\colhead{True ARs ($Y=1$)} &
\colhead{Ghost ARs ($Y=0$)}
}
\startdata
 Precision & 0.93 & 0.44 \\
 Recall    & 0.52 & 0.90 \\
 F1-score  & 0.67 & 0.59 \\
 Support & 1437 & 609 \\
\enddata
\end{deluxetable}

Diagnostic metrics summarized in \autoref{tab:model_performance} indicate that the model successfully rejects the majority of ghost \acp{ar} while maintaining a high precision of 0.93 for true \acp{ar}. However, this high precision comes at the cost of missing approximately 48\% ($1-$recall) of true \acp{ar} (see \autoref{tab:model_performance}), which is an expected consequence of adopting $P_{\rm th}$ specifically to minimize false detections. This trade-off is also evident in \autoref{fig:logistic_performence}a, which shows the probability distributions of the two classes. Although true \acp{ar} are generally assigned higher probabilities and ghost \acp{ar} are mainly concentrated toward lower probabilities, the substantial overlap between the two distributions implies that increasing $P_{\rm th}$ to achieve higher precision will always involve a trade-off between precision and recall, i.e., the ability to retain true \acp{ar}. In \autoref{fig:logistic_performence}b, we show the cumulative distribution function (CDF) of true \acp{ar} and the complementary CDF ($\mathrm{CCDF}:= 1.0-\mathrm{CDF}$) of ghost \acp{ar} as a function of the predicted $P$. These curves quantify how the choice of $P_{\rm th}$ affects the fraction of true \acp{ar} retained and the fraction of ghost \acp{ar} rejected, further illustrating the trade-off between recall and precision.

The resulting \acl{roc} (\acs{roc}, \autoref{fig:logistic_performence}c) yields an \acl{auc}, $\mathrm{AUC} = 0.77$, i.e., \acl{rocss}, $\mathrm{ROCSS}:= 2\mathrm{AUC}-1 = 0.54$, demonstrating discriminatory power between the two classes. The precision--recall relationship (\autoref{fig:logistic_performence}d) indicates a higher recall can be achieved without a substantial loss of precision, however tightening the target FPR can significantly change the recall. Finally, the resulting outcome matrix (\autoref{fig:logistic_performence}e) includes the relative fraction of each class based on $P_{\rm th}=0.73$. Collectively, these diagnostics demonstrate that the logistic \pmodel \ provides a quantitatively-evaluated and interpretable approach for identifying the true far-side \acp{ar} and its utility, despite the inherent uncertainties and complexities of helioseismic observations. 

We find that $\beta$ for all three physical parameters is positive and statistically significant with $p\text{-value} < 0.01$ (see \autoref{tab:predictor_summary}, $p\text{-values}$ are not shown). However, $\phihifar$ emerged as the dominant predictor, having a coefficient nearly twice that of $\bavg$ and an order of magnitude greater than that of $\cos\lambda$. This result is consistent with our physical understanding that stronger and more coherent magnetic regions produce more reliable helioseismic signatures, while latitude provides only a secondary constraint.

\subsection{Scaling Relationship Between $\Phi_\mathrm{FS}$ and $\Phi_{\rm HIFARM}$}

\begin{figure}[h!]
    \centering
    \includegraphics[width=\columnwidth]{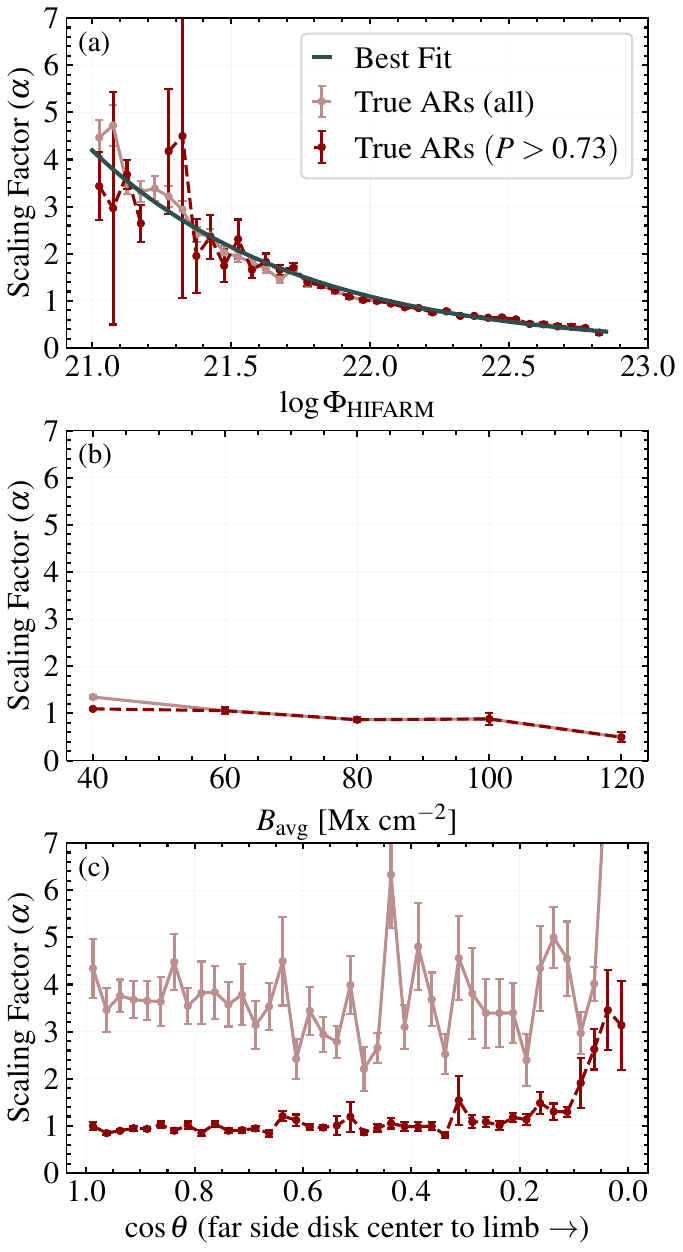}
    \caption{Variation of the flux scaling factor, $\alpha$ (\autoref{eq:alpha_flux}), computed at each instance of \ac{ar} detection. Two cases are considered: (i) all true \acp{ar}, and (ii) \acp{ar} with $P>0.73$. The mean of scaling factor, $\alpha$, is shown as a function of (a) $\phihifar$, computed in uniform logarithmic bins, (b) $\bavg$ and, (c) absolute $\cos\theta$. Error bars represent the corresponding \acl{sse}.}
\label{fig:flux_scaling}
\end{figure}

We show the variation of the mean $\alpha$ as a function of $\phihifar$, $\bavg$ and $\cos\theta$, in \autoref{fig:flux_scaling}a, \ref{fig:flux_scaling}b and \ref{fig:flux_scaling}c respectively. The mean value of $\alpha$ is calculated in uniform logarithmic bins of $\phihifar$ and uniform linear bins of $\bavg$ and $\cos\theta$. We examine this relationship in two cases, (i) using all true \acp{ar}, and (ii) using only those \acp{ar} for which the \pmodel\ predicts $P>0.73$. The latter corresponds to approximately 52\% of the true \acp{ar} population lying to the right of $P_{\rm th}=0.73$ (see \autoref{fig:logistic_performence}a, b).

We find that $\alpha$ decreases exponentially with increasing $\log(\phihifar)$ in both cases, and no statistically significant difference between the two populations. The primary difference is the larger error bars in case (ii), particularly at low $\phihifar$, which is expected because the stringent probability threshold substantially reduces the number of \acp{ar} in the low-flux bins. In contrast, the dependence of $\alpha$ on $\bavg$ (\autoref{fig:flux_scaling}b) is comparatively weak. We see a slight downward trend in $\alpha$ with increasing $\bavg$, but the magnitude of this variation is small compared to the dependence on $\phihifar$.We also find an increase in $\alpha$ near the far-side solar limb ($|\cos\theta|<0.1$), accompanied by larger uncertainties. This behavior is consistent with the fact that helioseismic detections near the limb have less sensitivity due to wave propagation geometries, leading to larger uncertainties in the inferred magnetic flux \citep{Zhao2019}. There is little to no dependence on $\cos\theta$ for values larger than $0.1$.  The lower overall values of $\alpha$ in \autoref{fig:flux_scaling}c for case (ii) are consistent with \autoref{fig:flux_scaling}a, as these \acp{ar} are associated with higher fluxes and therefore smaller scaling factors.

These results suggest that $\phihifar$ is the dominant parameter controlling the flux scaling relationship, whereas the dependence on $\bavg$ and $\cos\theta$ is comparatively weak,  especially for \acp{ar} meeting the $P>0.73$ criterion. Therefore, given the additional uncertainties associated with the far side helioseismic measurements, we neglect the contributions of $\bavg$ and $\cos\theta$ and model the scaling factor, $\alpha$, solely as a function of $\phihifar$ for all \acp{ar} with $|\cos\theta|<0.1$,
\begin{align}
    \log\alpha &= a\log(\phihifar)+b, \nonumber\\
    \text{i.e.,}\quad
    \alpha &= 10^b\,\left(\phihifar\right)^a,
    \label{eq:alpha_flux}
\end{align}
where $a$ and $b$ are free parameters. We obtain $a=-0.58\pm0.01$ and $b=12.87\pm0.24$ using the least-squares fitting algorithm. The negative power-law highlights that \acp{ar} with lower $\phihifar$ require proportionally larger flux corrections, whereas stronger counterparts have better flux estimates in \fsmag\ data.

\subsection{AFT Simulation with HIFARM}
\label{sec:far2aft}

We use the \pmodel\ (\autoref{eq:regressionmodel}) to calculate $P$ for all \acp{ar} detected in \fsmag, and create a catalog of \acp{ar} with $P>0.73$, i.e., true \acp{ar}.  The catalog contains the time of \fsmag\ maps, heliographic Carrington location ($\lambda, \phi$) and $\phihifar$ along with the timestamp of \fsmag,  ready to be ingested into the \ac{aft} simulation. The following steps describe the procedure for incorporating \fsmag-detected \acp{ar} into \ac{aft} based on the physical parameters extracted from the \fsmag.

\subsubsection{Creation of an Idealized Bipolar Magnetic Region}

To incorporate the selected \ac{ar} into \ac{aft}, we first construct an idealized Bipolar Magnetic Region (BMR) using the physical parameters derived from the \fsmag\ catalog. We begin by calculating $\alpha$, using \autoref{eq:alpha_flux}, to obtain the corrected far-side magnetic flux, $\fluxfs = \alpha \phihifar$. The area of an \ac{ar}, $A$ (in MSH, millionths of a solar hemisphere) is then estimated using the empirical relation,
$A = \fluxfs/(7.0\times10^{19})$,
where the magnetic flux is presented in Maxwells \citep{Sheeley1966, Mosher1977}.

Once the area is determined, the longitudinal separation between the two magnetic polarities is calculated as 
$\Delta\phi\degr = 3\degr + 8\degr\tanh{\left(A/500.0\right)}$,
following \citet{Hathaway2016, Upton2024}. The corresponding latitudinal separation is computed using the relation 
$\Delta\lambda\degr = \gamma \Delta\phi$,
where $\gamma$ is the BMR tilt angle relative to the solar east--west direction \citep{Hale1919, Howard1991a}. We estimate the tilt using Joy's law, $ \gamma = 32.1\degr\sin\lambda$,
established from previous observational studies \citep{Stenflo2012a, Jha2020, Sreedevi2023}. Using these quantities, the physical parameters $(\lambda, \phi, \fluxfs)$ of the two BMR polarities are given as
$$
\left(\lambda^\pm, \phi^\pm, \fluxfs^\pm\right)
=
\left(
\lambda \pm \frac{\Delta\lambda}{2},
\phi \pm \frac{\Delta\phi}{2},
\pm \frac{\fluxfs}{2}
\right).
$$

Finally, each polarity is represented using a two-dimensional Gaussian magnetic field distribution, together forming an idealized BMR ready to be incorporated into the \ac{aft} simulation.

\subsubsection{Incorporating ARs in AFT Model} \label{subsec:aft_hifar}

Taking into account the uncertainties and assumptions involved in constructing BMRs from the far-side from the limited physical information available in the \hmifs\ data, we incorporate true \acp{ar} into \ac{aft} using the following procedure.

At a given time step, $t$, we again consider a rectangular region $[\lambda\pm5\degr,\phi\pm10\deg)$ and calculate the total unsigned magnetic flux from the simulated \ac{aft} magnetic field map, $\Phi_{\rm AFT}(t)$. The incorporation of each \ac{ar} is then determined individually according to the following criteria:

\begin{enumerate}[left=-10pt, label={(\textit{\roman*})}]
    \item If $\Phi_{\rm AFT}(t) \ge \fluxfs(t)$, the \ac{ar} is not incorporated under the assumption that the region is either already sufficiently represented in the simulation or is in a decaying phase. This criterion prevents the introduction of spurious magnetic flux into the model.

    \item If $\Phi_{\rm AFT}(t) < \fluxfs (t)$, we calculate the fraction of missing flux as 
    $$f(t) = \frac{\fluxfs(t) - \Phi_{\rm AFT}(t)}{\fluxfs (t)},$$
    and scale the idealized BMR associated with the \ac{ar} by the factor $f(t)$. 

    \item The scaled idealized BMR is added to AFT. This approach ensures that only the missing magnetic flux is incorporated into \ac{aft} at each time step.
\end{enumerate}

 In Appendix~\ref{appendix:bfly}, we present butterfly diagrams showing the latitude–time distribution of far-side \acp{ar} at each stage of the filtering and selection process.

\begin{figure}[h!tbp]
    \centering
    \includegraphics[width=\columnwidth]{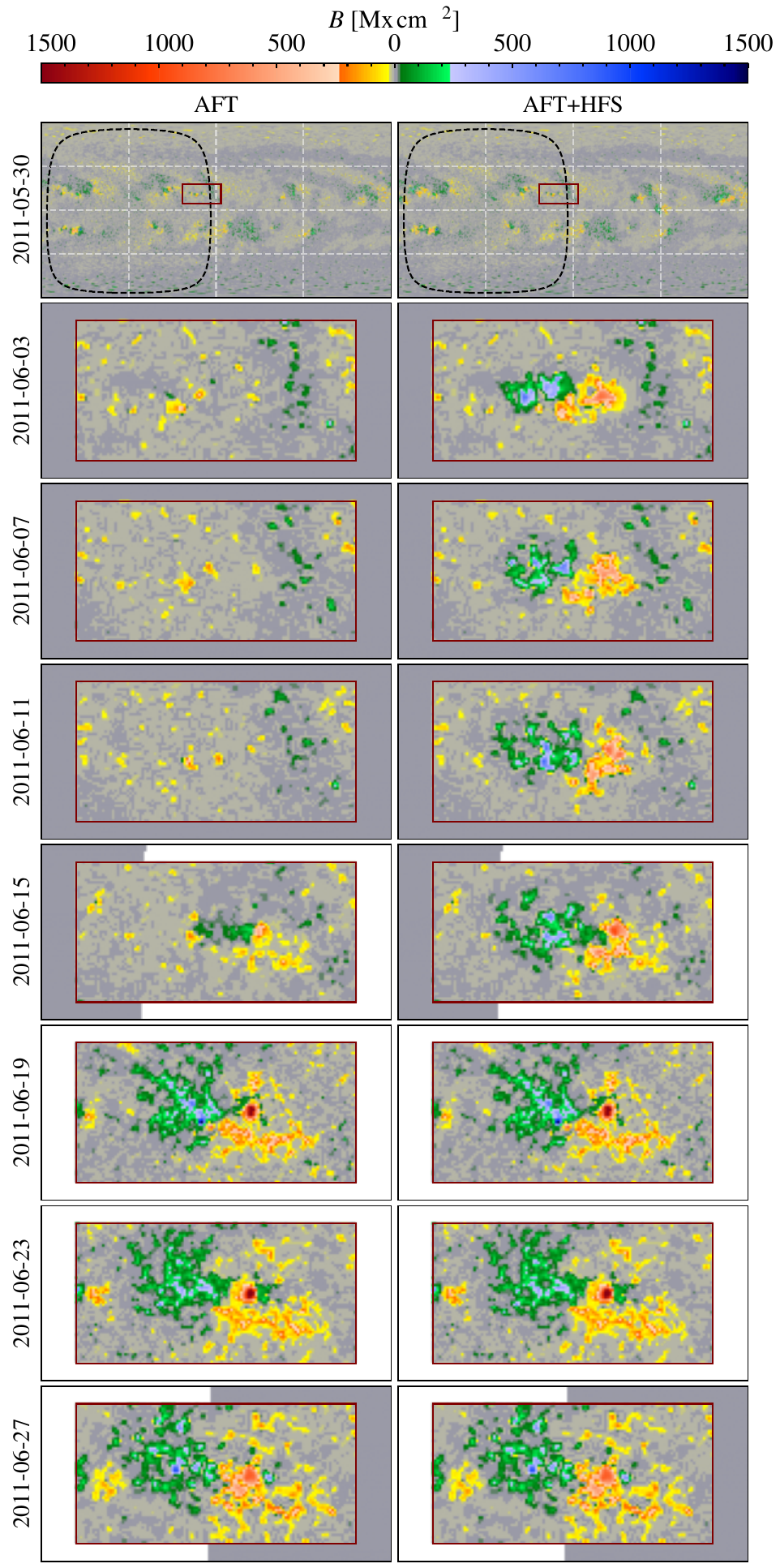}
    \caption{Representative example of the evolution of an \ac{ar} (red rectangular box), \texttt{TDL 20110530TN17165} (NOAA AR11236; see text), that emerged near the near-side west limb of the Sun on 2011 May 30 (at Carrington coordinates N17, 165), comparing the \ac{aft} baseline simulation (left column) with the far-side-informed \ac{aft} (\afthfs) simulation (right column) from 2011 May 30 to 2011 June 27. Dashed black contours in the top row indicate the data-assimilation window used by \ac{aft}, whereas vertical and horizontal white dashed lines are reference grid lines. The white and gray regions in the remaining rows denote the near-side and far-side hemispheres, respectively, illustrating the evolution of the \ac{ar} before and after its rotation onto the visible hemisphere.}
    \label{fig:afthifar_example}
\end{figure}

\autoref{fig:afthifar_example} demonstrates the evolution of an \ac{ar} in the \ac{aft} baseline and \afthfs\ simulations. We observe an early signature of the emergence of an \ac{ar} near the western limb of the near-side hemisphere around the 2011 May 30. As the region rotates onto the far side of the Sun, \ac{ar} emergence does not continue in the \ac{aft} baseline simulation, as expected. However, \fsmag\ detects the region with $P>0.73$, and by incorporating this detection into \ac{aft}, the model continues to follow the growth and evolution of the \ac{ar} during its far-side passage. This \ac{ar} reappears on the visible hemisphere on 2011 June 15, and merges smoothly with near-side observations as data assimilation takes over. This highlights that despite implementing an idealized bipolar \ac{ar} on the far side, the convective flows included in \ac{aft} \citep{Upton2014, Upton2014a} quickly disperse the magnetic flux concentration in the bipolar region and evolve it into a realistic magnetic-field distribution, evident from smooth transition of the \ac{ar} from far-side to near-side (\autoref{fig:afthifar_example}). 

\begin{figure}[!h]
    \centering
    \includegraphics[width=\columnwidth]{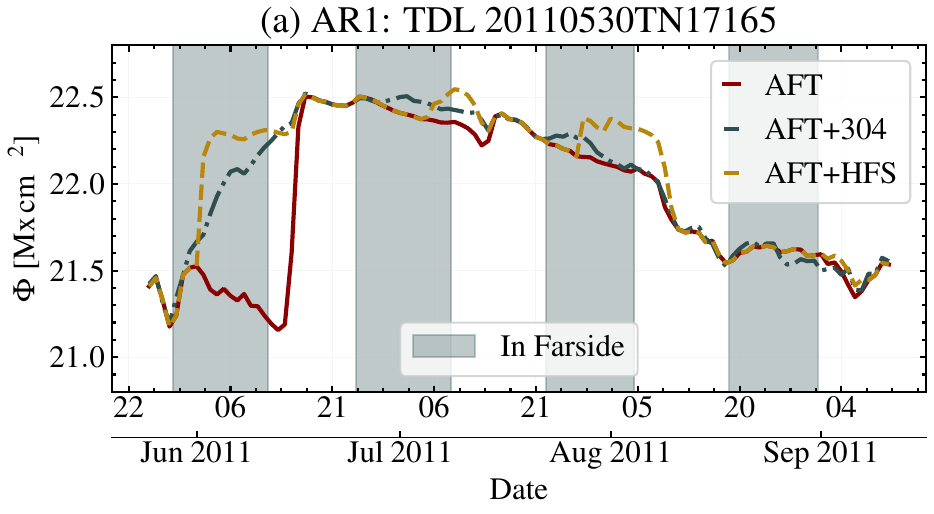} 
    \includegraphics[width=\columnwidth]{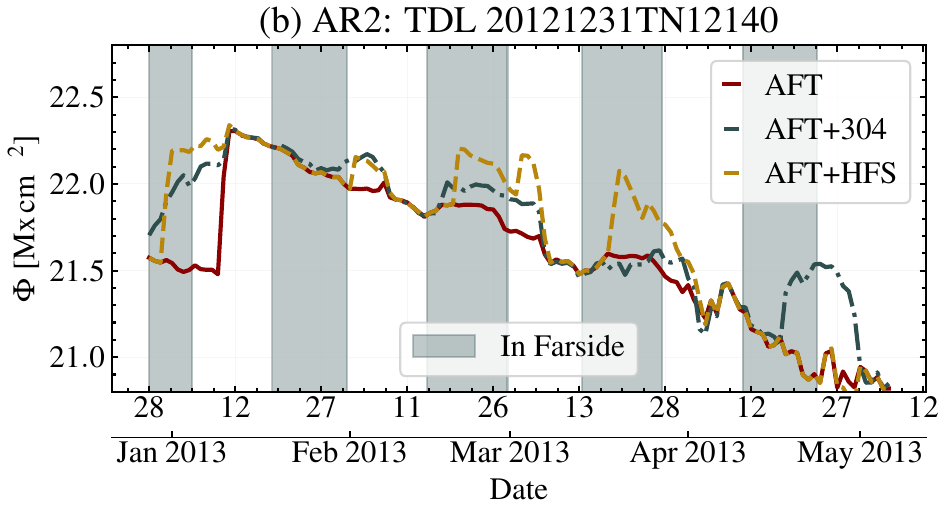} 
    \caption{Evolution of the total unsigned magnetic flux of two representative \acp{ar} in the \ac{aft} baseline, \afthfs, and AFT+304 simulations. (a) \texttt{TDL~20110530TN17165} appeared on 2011 May 30 near the near-side western limb, subsequently evolved on the far side of the Sun and reappeared at the eastern limb on 2011 June 15 as NOAA AR11236. (b) \texttt{TDL~20121231TN12140} first detected in \fsmag\ on 2012 December 31 and subsequently observed near the eastern limb after 2013 January 06 as NOAA AR11652. The shaded regions in both panels indicate the period during which the \acp{ar} were located on the far side of the Sun.}
    \label{fig:arflux}
\end{figure}

We show the evolution of the total unsigned magnetic flux of two representative \acp{ar}, (AR1) \texttt{TDL~20110530TN17165} (NOAA 11236, \autoref{fig:arflux}a) and (AR2) \texttt{TDL~20121228TN12140} (NOAA AR11652, \autoref{fig:arflux}b), as they complete multiple rotations across the near-side and far-side hemispheres of the Sun. For ``TDL'' (time, date, location) \ac{ar}-identification scheme see \citet{Leka2026}. For each region, we compute the total unsigned flux from the \afthfs, \ac{aft} baseline, and AFT+304 photospheric magnetic field maps and compare their temporal evolution.

AR1 and AR2 both feature significant far-side emergence that is not captured in the AFT Baseline, but is included in AFT+304 and \afthfs. The flux evolution for AR1 (\autoref{fig:arflux}a) in \afthfs\ remains well aligned with AFT+304 throughout the period shown, whereas in case of AR2 (\autoref{fig:arflux}b), we observe some notable differences between the two simulations. In particular, \afthfs\ produces higher flux during 2013 March 13\,--\,March 28 and lower flux during 2013 April 12\,--\,April 27, compared to AFT+304. In both AFT+304 and \afthfs, such differences appear to deviate from the true \ac{ar} evolution and are corrected by the data assimilation of the near-side observations. These discrepancies may arise from uncertainties in the magnetic flux inferred from \euvi\ intensities through the flux--luminosity relationship used in AFT+304 \citep{Upton2024}, or from uncertainties in $\alpha$, used to convert $\phihifar$ into the flux incorporated in \afthfs\ (in this work). These cases demonstrate that while these data products are useful for detecting far-side \ac{ar} emergence and creating simulations with more accuracy on the fer side, regular direct magnetic observations are needed to correct for any excess \ac{ar} flux inferred by either of these data products.

\begin{figure}[!h]
    \centering
    \includegraphics[width=\columnwidth]{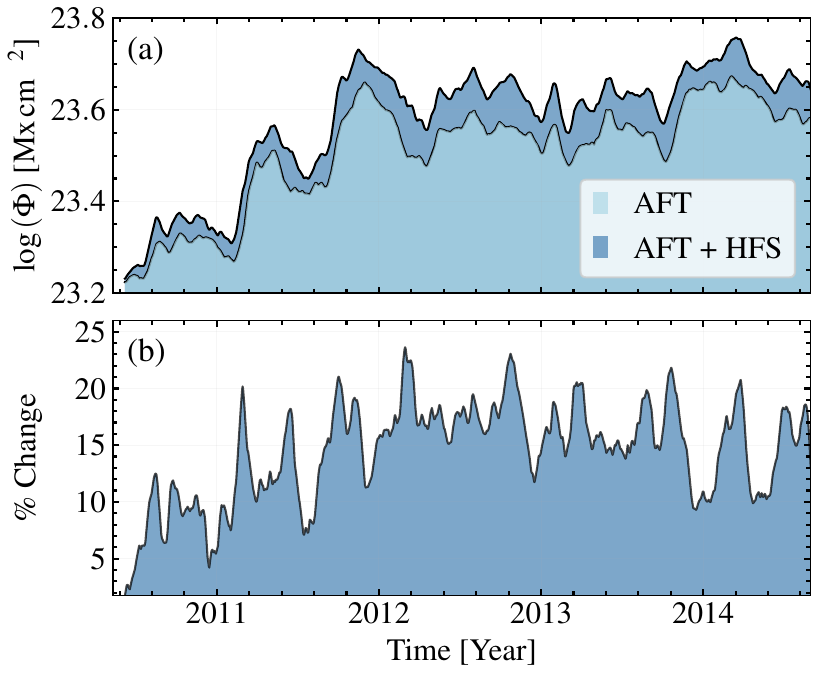}
    \caption{Global unsigned magnetic flux in the \ac{aft} baseline and \afthfs\  simulations. (a) $\Phi$, integrated over the solar surface. (b) Percentage flux difference of the \ac{aft} baseline relative to the \afthfs\ simulation.}
    \label{fig:globalflux}
\end{figure}

To quantify the contribution of far-side \acp{ar} to the global magnetic flux budget of the Sun, defined here as the total unsigned magnetic flux integrated over the solar surface, we compare the \ac{aft} baseline and \afthfs\ simulations. \autoref{fig:globalflux}a shows the variation of total unsigned flux in the two simulations, while \autoref{fig:globalflux}b shows the percentage flux deficit in the \ac{aft} baseline relative to \afthfs. These results demonstrate that far-side \ac{ar} flux makes a significant contribution to the global magnetic flux budget and that incorporating \fsmag\ detections reduces the loss of magnetic flux that is otherwise missed in \ac{sft} simulations driven solely by near-side observations. \citet{Virtanen2021} also demonstrated that over longer timescales, far-side \acp{ar} may influence the evolution of the polar field using the simulated far-side \acp{ar}. However, this effect may depend on the data assimilation approach in the \ac{sft} model, and quantifying it requires a longer data set, which is left for a follow-up investigation.

\section{Conclusion}\label{sec:conclusion}
A major limitation of most existing data-driven \ac{sft} models is their inability to account for \acp{ar} that emerge and grow on the far side of the Sun. In this work, we address this limitation by developing and validating a probabilistic framework for detecting and incorporating helioseismic far-side \acp{ar} from \fsmag\ maps into the \ac{aft} model. 

Taking advantage of the near-simultaneous data from \euvi\ and \fsmag, together with the AFT+304 simulations for the period from 2010 May 13 to 2014 August 18, when \ac{stereo} provided substantial coverage of the solar far side. We demonstrate that the \pmodel\ developed in this work is very successfully at distinguishing true \acp{ar} from ghost \acp{ar} in \fsmag\ data. The analysis shows that the total magnetic flux is the dominant physical parameter of a true \ac{ar}, while the average magnetic field strength and latitude provide additional discriminatory power.  The \pmodel\ developed in this work is not specific to \ac{aft} and can be easily integrated into other \ac{sft} models, such as \ac{adapt} -- mostly used currently, \acl{hipft} \citep{Caplan2025}, \acl{farm}\citep{Yang2024} and others.

We find that the magnetic flux inferred from \fsmag\ maps requires a systematic correction by $\alpha$, which arguably depends primarily on $\phihifar$ (\autoref{eq:alpha_flux}), before incorporation into \ac{aft}. This scaling is essential to obtain the physically consistent amount of magnetic flux for the model. As a proof of concept, with examples, we demonstrate that this framework reproduces the flux evolution of \acp{ar} in good agreement with AFT+304 during their far-side passage across multiple solar rotations. At the global scale, we find that the \ac{aft} baseline simulation underestimates the total unsigned magnetic flux by approximately 10--20\%, demonstrating that the far-side \acp{ar} make a significant contribution to the global magnetic flux budget and the evolution of the full-Sun photospheric magnetic field. However, by adopting a higher $P_\mathrm{th}$, we also exclude some true \acp{ar} (false negatives), which may contribute to the total magnetic flux budget. Although the missed true \acp{ar} with lower $P$ are associated with relatively low magnetic flux, improving the classification model, e.g., by adding helioseismic sensitivity as a function of $(\lambda, \phi)$, higher moments of magnetic filed and flux distribution, measures of spatial compactness and fragmentation, prior information of near-side \acp{ar}, etc. could further reduce these false negatives while maintaining good discrimination.

The ability of the \afthfs\ framework to capture the emergence and evolution of \acp{ar} throughout the entire solar surface enables a more complete reconstruction of the full-Sun photospheric magnetic field. These more accurate magnetic field maps thus provide improved inner boundary conditions for coronal and heliospheric models in studies of the magnetic coupling between the solar surface, corona, and solar wind. Furthermore, the current implementation uses a single \fsmag\ map to estimate the probability of true \acp{ar}, making it ideal for ensemble modeling with quantified uncertainties.  This framework provides a practical pathway for incorporating helioseismic far-side observations into data-driven models used in operational space-weather forecasting, including full-Sun ($4\pi$) flare forecasting frameworks \citep[e.g.,][]{Leka2026}.

In a series of studies, \citet{Hamada2024, Hamada2025} developed \acl{ml}-based methods that, similar to this work, combine \euv\ observations with far-side phase-shift maps (from \ac{gong}) to distinguish true \acp{ar} from ghost \acp{ar}. In addition, \citet{Hamada2026} proposed a method for identifying the magnetic polarities of far-side \acp{ar}, providing additional morphological constraints for constructing bipolar magnetic regions. The versatile design of the \afthfs\ framework makes it straightforward to incorporate both their classification and polarity separation methods as an extension. Integrating these complementary approaches would enable a systematic assessment of the impact of far-side \acp{ar} on the reconstruction of the far-side magnetic field while also simplifying the space-weather forecasting pipeline through the direct incorporation of \ac{gong} helioseismic phase-shift maps.

Beyond its application to far-side \acp{ar} detection, the \pmodel\ approach provides a physically interpretable diagnostic method for investigating the physical processes that govern the detectability of helioseismic far-side \acp{ar}. By explicitly relating the detection probability to observable properties of active regions and their viewing geometry, the model enables a quantitative assessment of the factors that limit helioseismic far-side detection. Such physical insight not only improves our understanding of the strengths and limitations of current helioseismic techniques but also provides a foundation for guiding future developments in far-side helioseismic observations, detection methods, and instrument design.


\begin{acknowledgments}
BKJ, LAU, KDL and KJ were partially supported by NASA/R2O2R 80NSSC22K0273. BKJ, LAU, RC, JZ, SAHW and KJ were partially supported by NASA DRIVE Science Center COFFIES Phase II CAN 80NSSC22M0162. RC, JZ, and SAHW were partially supported by the NASA SDO/HMI contract NAS5-02139, relating to HSO support. BKJ and LAU acknowledge support from NASA Heliophysics Living With a Star grant 80NSSC23K0048 and KDL was further supported by NASA LWS/SC 80NSSC22K0893. IUU acknowledges support from the Office of Naval Research. HMI data used in this study are courtesy of NASA-SDO and the HMI instrument team. This research has made use of NASA's Astrophysics Data System (ADS; \url{https://ui.adsabs.harvard. edu/}) Bibliographic Services. 
\end{acknowledgments}

\begin{contribution}
LAU developed the \ac{aft} model. BKJ and LAU jointly developed the \pmodel\ and the \afthfs\ framework. KDL, JZ and KJ provided key scientific and technical input that contributed to the development of the \afthfs\ framework. RC, JZ and SAHW provided the \fsmag\ dataset used in this study and support thereof. IU provided the full-Sun SDO/AIA and STEREO/EUVI 304\,\AA\ composite observations used to construct the labeled \ac{ar} catalog and to generate the AFT+304 simulations (created by LAU and IUU). BKJ prepared the initial draft of the manuscript with contributions from LAU. All authors provided comments and revisions that improved the final manuscript.


\end{contribution}

%
\facilities{SDO (HMI), SDO (AIA), STEREO (EUVI)}

\software{Matplotlib \citep{Hunter2007}, Numpy \citep{numpy}, SunPy\citep{sunpy}, Pandas \citep{pandas}, SciPy \citep{scipy}, Astropy \citep{astropy}, statsmodels\citep{seabold2010} and AFTpy \citep{aftpy}}

\appendix
\section{Comparison of Logistic Regression Models}\label{appendix:comparision}

The logistic \pmodel{s} considered during predictor selection are summarized in \autoref{tab:model_comparison}. All models were fitted using maximum-likelihood logistic regression to the same set of 8181 observations. Each model includes $\cos\lambda$ and the total unsigned magnetic flux, $\phihifar$ and additional parameters are added incrementally to evaluate their contribution to the model fit. Model performance is quantified using the maximized log-likelihood, $\ln\mathcal{L}$, and McFadden's pseudo-$R^2$, with $\Delta R^2$ representing the increase relative to the preceding model. Although including $B_{\rm max}$ shows small improvement, it was excluded from the adopted model for physical interoperability ($B_{\rm max}$ is more prone to helioseismic artifacts). 

\begin{deluxetable}{llcccl}[!htbp]
\tablecaption{Comparison of logistic \pmodel{s} considered for \hmifs\ \ac{ar} identification. $\ln\mathcal{L}$ is the maximized log-likelihood of the fitted model, and $\Delta R^2$ denotes the increase in pseudo-$R^2$ relative to the preceding model. Standardized coefficients are listed in the same order as the physical parameter ($\mathbf{x}$).\label{tab:model_comparison}}
\tablewidth{0pt}
\tablehead{
\colhead{Model} &
\colhead{Physical Quantities $(\mathbf{x})$~~~} &
\colhead{$\ln\mathcal{L}$} &
\colhead{Pseudo-$R^2$} &
\colhead{$\Delta R^2$} &
\colhead{Standardized Coefficients $(\mathbf{\beta})$~~~~}
}
\startdata
M1 & $\cos\lambda,\ \phihifar,\ B_{\rm max}$ & $-4286.4$ & 0.1472 & \nodata & 0.079,\ 1.777,\ 0.066 \\
M2 & $\cos\lambda,\ \phihifar,\ B_{\rm avg}$ & $-4229.4$ & 0.1586 & 0.0114 & 0.066, \ 1.225,\ 0.640 \\
M3 & $\cos\lambda,\ \phihifar,\ B_{\rm avg},\ B_{\rm max}$ & $-4179.9$ & 0.1684 & 0.0098 & 0.061,\ 1.504,\ 1.152,\ -0.963 \\
M4 & $\cos\lambda,\ \phihifar,\ B_{\rm avg},\ B_{\rm max},\ \cos\theta$ & $-4179.8$ & 0.1685 & 0.0001 &  0.060,\ 1.498,\ 1.153,\ -0.960,\ 0.014
\enddata
\end{deluxetable}

\section{HIFARM Butterfly Diagram}\label{appendix:bfly}

\begin{figure}[!htbp]
    \centering
    \includegraphics[width=\textwidth]{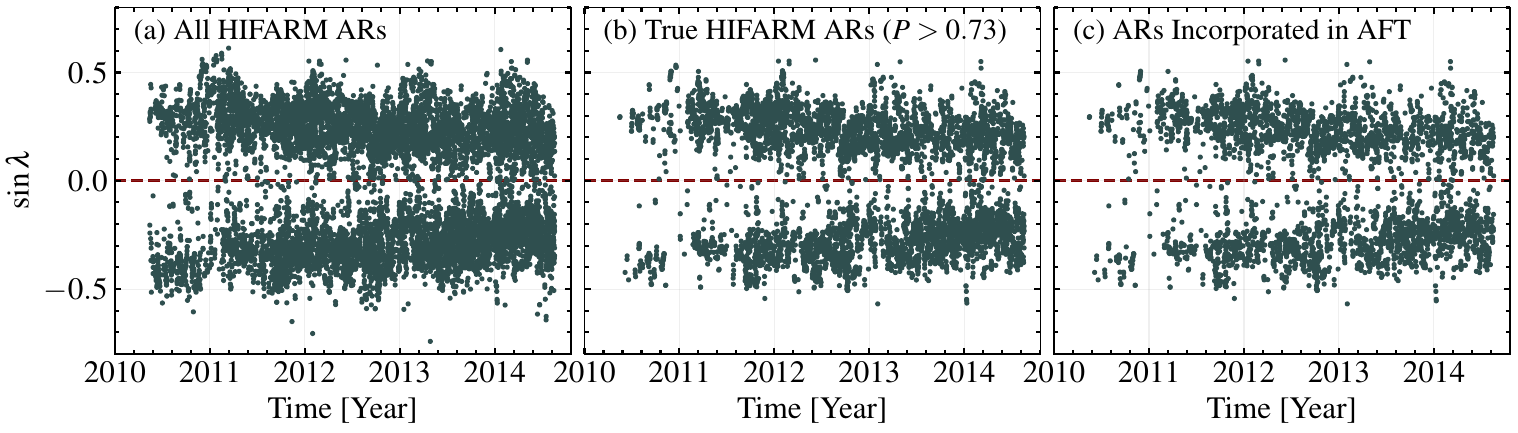}
    \caption{$\sin\lambda$ of each instance of \acp{ar} as a function of time (the butterfly diagram) for (a) all candidate \acp{ar} identified in \fsmag, see Section~\ref{subsec:ar_detection}, (b) \acp{ar} with $P \ge 0.73$, and (c) \acp{ar} incorporated into \ac{aft} based on the criteria discussed in Section~\ref{subsec:aft_hifar}.}
    \label{fig:hifar_butterfly}
\end{figure}

\autoref{fig:hifar_butterfly} presents butterfly diagrams showing $\sin\lambda$ of the detected \acp{ar} as a function of time at three stages of the \afthfs\  framework. \autoref{fig:hifar_butterfly}a shows all candidate \acp{ar} identified by the \ac{ar} detection algorithm described in Section~\ref{subsec:ar_detection}. \autoref{fig:hifar_butterfly}b includes only \acp{ar} with $P \ge P_\mathrm{th}$ of $0.73$. A similar butterfly diagram is shown \citet[Figure~8]{Hamada2024} based on the \ac{gong} far-side helioseismic pipeline. Finally, \autoref{fig:hifar_butterfly}c shows the subset of \acp{ar} incorporated into the \afthfs\ simulation based on the criteria described in Section~\ref{subsec:aft_hifar}. The difference between \autoref{fig:hifar_butterfly}b and \autoref{fig:hifar_butterfly}c is that \acp{ar} already sufficiently represented in \ac{aft} ($\Phi_{\rm AFT}(t) \ge \fluxfs$) are not incorporated. Most of these excluded \acp{ar} are either in the decaying phase or have already been incorporated with sufficient magnetic flux.

\bibliography{formated_all}{}
\bibliographystyle{aasjournalv7.1}
\end{document}